# Beyond the Desktop: Emerging Technologies for Supporting 3D Collaborative Teams

Jannick Rolland,[1-2] Ozan Cakmakci,[1] Jeff Covelli,[2] Cali Fidopiastis,[1-2] Florian Fournier,[1] Ricardo Martins,[2] Felix Hamza-Lup,[3] and Denise Nicholson[2]

(1) : CREOL, College of Optics & Photonics, University of Central Florida, 4000 Central Florida Blvd., Orlando, Florida, 32816
+1.407.823.6870/+1.407.823.6880
E-mail : {jannick, ozan, florian}@odalab.ucf.edu

(2) : Institute for Simulation and Training, University of Central Florida, 3100 Technology Parkway, Suite 319, Orlando, Florida, 32826
+1.407.882.1444/+1.407.882.1335
Email :{jcovelli,cfidopia,ricardo,dnichols}@ist.ucf.edu

(3) : Computer Science, Armstrong Atlantic State University, 11935 Abercorn Street
Savannah, Georgia 31419
+1.912.921.5668
E-mail : { felix}@cs.armstrong.edu

**Abstract:** The emergence of several trends, including the increased availability of wireless networks, miniaturization of electronics and sensing technologies, and novel input and output devices, is creating a demand for integrated, fulltime displays for use across a wide range of applications, including collaborative environments. In this paper, we present and discuss emerging visualization methods we are developing particularly as they relate to deployable displays and displays worn on the body to support mobile users.

## 1. Introduction

The design of head-worn displays (HWDs) is complex being driven by ergonomics, information content and tasks performance, and dominated by brightness and weight concerns.[RH1] The design of ultra-compact optics approaching the eyeglass form factor is important for example in applications such as mobile information displays.[CR1] On the other hand, if the task is flying a commercial airliner or training a military pilot on a new or developing flight platform full immersion provided by a wide field of view HWDs of the order of 80 deg. per eye is proving to be necessary [CR2, R1] The emerging displays being developed right now at the Optical Diagnostics and Applications Laboratory (ODALab) are see-through and look-around HWDs, with a focus on projection type displays and off-axis designs.[CR3, R2] To-date, displays have been mainly developed around look-at technologies, which place too much weight cantilevered on the face and when combined with the dangers of being cutoff from the surround have not been adopted by the market. Off-axis, non-rotationally symmetric designs have been mainly avoided due to a lack of optical design methods and lagging fabrication capabilities for nonsymmetric surface shapes. The state-of-the-art display modalities emerging from ODALab are capitalizing on new optical design methods developed under DARPA funding combined with the new to the market 6-axis diamond turning machines. Perhaps most enabling is new LED-based illuminators and generally brighter display technologies are allowing the displays to move out into to full sun with fully user-accepted performance.

## 2. Featured Technologies

In Figure 1, we feature (on the left) our latest projection HWD, with a projection screen integrated within the HWD that enables outdoor applications. The projection screen is made of retroreflective material to maximize the amount of light collected by the user's eyes. In this system, ultra-compact optics mounted on a stretchable hat for uniform distribution of the weight on the user's head allows imaging of the projection screen in front of the user, as if it was physically there.[RM1] On the right, we feature our first EyeGlass HWD that uses only two optical elements, an off-axis free form mirror and a hybrid refractive/diffractive lens.[CR3] We are using the term free-form in reference to surfaces that are non-rotationally symmetric such as x-y or Zernike polynomials. Advances in optical fabrication are enabling non-rotationally symmetric free-form surfaces at price points that are not market prohibitive.

### 3. Alternative Approaches

In the past, the limitation of on-axis optical design forms has led to two particular paths when attempting the design of eyeglass form HWDs. One approach, to minimize weight is to use holographic optical elements (HOE). To-date, the challenge has been to design a full color (red, green and blue) display based on an HOE. A head-worn, look-at display with a 3mm exit pupil, 27x10 degrees of field of view, operating at the single wavelength of 532nm was recently designed based on a HOE and fabricated by Minolta.[KT1] A radical alternative in the design of HWDs is to use a laser source to temporally scan the image on the retina without intermediary image formation. The concept is in principle attractive to yield high brightness required in various applications. The challenging factor here is the Lagrange invariant combined with high speed scanning which results in a requirement that the product of the scan angle to the scanner lateral extent be constant throughout the optical system. Because of the high speed scanning, the scan angle is typically limited to less than 10 degrees and a larger scan angle at the eye that portray the FOV necessarily imposes an equivalent demagnification of the pupil. As a consequence, the exit pupil in such HWD is on the order of 1-2mm and pupil expansion mechanisms are required. Pupil expanders add additional complexity and size to the system and also reduce perceived brightness, creating in practice unique trade-offs. An alternative to higher luminance sources is to dim the light from the scene using polarizers, or yet more desirable electrochromic or photochromic mechanisms. [RM1, K1, MD1].

### 4. Design Space and Methods

Driving factors in the overall size of any HWDs are the microdisplay size and the eye-relief. The smaller the microdisplay, the shorter will be the focal length of the optics for the same FOV, and thus the most compact the HWD be. For the eyeglass form HWD, we find display sizes between 0.25" and 0.44" to be desirable. Naturally, the resolution of the microdisplay may be thought of as a number of dots per inch, thus often while not always, the resolution decreases with smaller sizes, which will strongly impact the decision process. The choice for eye relief is highly application-dependent, and large eye relief will allow adapting various corrective glasses often at the expense of compactness. However for systems above 500 grams, the most critical parameter shifts from the weight to the placement of the center of gravity of the overall system. Importantly, the level of comfort experienced by the user is ultimately the final judging factor. Designs in our laboratory have capitalized since 1998 on the use of plastic material as well as hybrid refractive-diffractive optics as they both yield lighter systems.[RH1] The EyeGlass HWD entire assembly weighs approximately 250 grams including the outer shell,

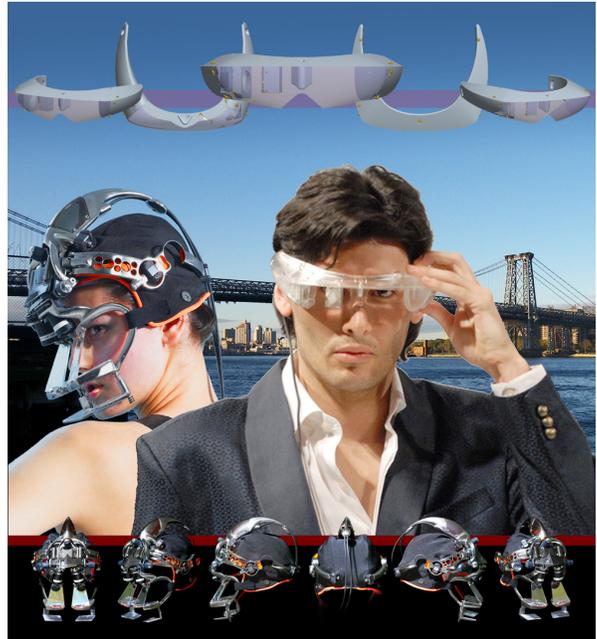

**Graphic Courtesy of Human Artifact R&D.**

**Figure 1:** Demonstration of the emergence of prototype see-through HWDs that will change how information will be displayed by the end of the decade in both military and consumer markets. On the left, Klaudia Zdanowicz of NEXT Model Management was photographed by Teddy Maki. On the Right, credit goes to Cheryl Delgreco of Media Strategies CTIA. The Optical System Designs were developed by the ODALab at the University of Central Florida and the Industrial and Optomechanical Designs by Human Artifact R&D.

optics, optomechanical mounts, microdisplay and the microdisplay driver electronics circuit board.

In HWDs, the pupil of the human eye is typically the aperture stop of the system. Therefore, the aperture stop and the exit pupil coincide in this system. It is customary to optimize HWDs across a finite pupil size that is significantly larger than that of the human eye to accommodate natural eye movements up to ±20deg and alignment of the eye to the HWD. Since under photopic illumination the pupil of the human eye is about 3mm, the visual performance analysis is often conducted for a 3mm eye pupil for both centered and decentered pupils within the overall eye box for which the system was originally optimized. The polychromatic MTF evaluated for a centered 3mm pupil for the on-axis field and the performance limiting fields and decenters serves to evaluate the amount of contrast that will be perceived as a function of the level of detail or spatial frequencies of the information being displayed. In HWDs, the overall system MTF is often limited by the pixel size, which sets the Nyquist or limiting spatial frequency for the optical system. Choices in the spatial frequency cutoff that are not conventional in designing optics may improve the subjective image quality perceived by users.

It is critical that all HWDs be quantatively benchmarked for spatial visual acuity, which will vary with target contrast as well as depth perception in the case of stereoscopic systems.[F1] For some applications, HWDs should also be benchmarked for temporal visual acuity and the perception of color, although these are difficult metrics to quantify through in-use measurement. In visual space, the EyeGlass HWD shown in Fig. 1 provides across a 20 degree FOV, 1.5 arc minutes resolution as limited by the pixel size in the microdisplay. The outdoor projection HWD provides 2.5 arc minute resolution across a 42 deg. FOV. . The human eye is limited to 1 arc minute resolution by the 2.5 μm cone spacing. It is possible however to resolve beyond 1 arc minute when accounting for neural processing.

## 5. Manufacturability Issues

In the design of any non-conventional optics, whether free form, HOE, or diffractive elements are used, a next critical component is manufacturability. Diffractive optical elements exhibit complementary dispersion to that of optical glasses and plastics and as a result they are extremely effective for weight reduction in a broadband display application. A question that often arises is whether such components can be manufactured to design specification and at what cost. First, custom optics elements cost many times, often more than a factor of 10, that of catalog optics. For a custom-designed, hybrid, refractive-diffractive element, the cost of that element is typically about twice that of a refractive element, but the reduction in weight that results can be dramatic. In the case of hybrid refractive-diffractive elements, diamond turning techniques can produce an aspheric substrate along with the diffractive optical element at no additional cost. [SG1,LY1] The most exciting, relevant emerging technology is that of free form optical surfaces produced by 6-axis diamond turning machines. These machines allow the optical designer to gain degrees of freedom previously unavailable to maintain performance while reducing weight.

## 6. Acknowledgements

This work was supported in part by the Florida Photonics Center of Excellence (FPCE), by the National Science Foundation (NSF) grant IIS/HCI 03-07189, and by the Office of Naval Research (ONR) Virtual Technologies and Environments (VIRTE) program. We thank Optical Research Associates for the educational license of CODEV® and Adam Oranchak for his expertise with optomechanics. Jannick Rolland email address is jannick@odalab.ucf.edu.

## 7. References

**[CR1]** Cakmakci O. and Rolland J.P. Head-Worn Displays: A Review. Journal of Display Technology 2(3):199-216, 2006.

**[CR2]** Covelli, J.M., Rolland J.P. and Hancock P. A Quantitative Measurement of Presence in Flight Simulators, Proceedings of ITSEC, 2007.

**[CR3]** Cakmakci O. and Rolland J.P. Design and fabrication of a dual-element off-axis near-eye optical magnifier. Optics Letters 32(11):1-3, 2007.

**[KT1]** Kasai I., Tanijiri, Y., Endo, T., and Ueda, H. A Forgettable Near-Eye Display. In Proc. Of the Fourth International Symposium on Wearable Computers 115-118, 2000.

**[MD1]** Mortimer R., A., Dyer L. and Reynolds J. R. Electrochromic organic and polymeric materials for display applications. Displays 27:2–18, 2006.

**[R1]** Rogers J. Design of an advanced helmet mounted display (AHMD), SPIE 5801:304-315, 2005.

**[R2]** Rolland J.P. Wide angle, off-axis, see-through head-mounted display. Optical Engineering - Special Issue on Pushing the Envelop in Optical Design Software, 39(7): 1760-1767, 2000.

**[RH1]** Rolland, J.P. and Hua, H. Head-mounted displays. in Encyclopedia of Optical Engineering, R. Barry Johnson and Ronald G. Driggers, Eds..,2005.

**[RM1]** Rolland J.P., Martins R. and Ha Y. Head-mounted display by integration of phase-conjugate material," US Patent 6,963,454, Nov 8, 2005.

**[F1]** Fidopiastis C. User-Centered Virtual Environment Assessment and Design for Cognitive Rehabilitation Applications. Ph.D. Disssertation, University of Central Florida, 2007.

**[SG1]** Stone T. and N. George. Hybrid diffractive-refractive lenses and achromats. Applied Optics 27(14): 2960-2971, 1998.

**[LY1]** Li L., Yi. A.Y. and Huang C. Fabrication of diffractive optics by use of slow tool servo diamond turning process. Optical Engineering 45(11): 3401-3410, 2006.